\documentclass[aps, prl, amsmath,amssymb,notitlepage, twocolumn, nofootinbib]{revtex4-1}

\usepackage{subfigure}
\usepackage{amssymb}
\usepackage{graphicx} 
\usepackage{color}
\usepackage{amsmath}
\usepackage[colorlinks=true,linkcolor=blue,citecolor=blue,urlcolor=blue]{hyperref}
\usepackage{framed}
\usepackage{gensymb}

\newcommand{\be}{\begin{equation}}
\newcommand{\ee}{\end{equation}}

\newcommand{\bra}[1]{ \left \langle #1 \right| }
\newcommand{\ket}[1]{ \left| #1 \right \rangle }
\newcommand{\Outer}[2]{ \left| {#1} \middle\rangle \middle\langle  {#2} \right| }
\newcommand{\trace}[1]{ \mbox{Tr} \left\{ {#1} \right\} }
\newcommand{\Trace}[2]{ \mbox{Tr}_{#1} \left\{ {#2} \right\} }
\newcommand{\subsc}[1]{ \mbox{\scriptsize{#1}} } 

\begin{document}

\title{ Metasurface imaging with entangled photons}

\author{Charles Altuzarra$^{1,+}$, Ashley Lyons$^{2,+}$, Guanghui Yuan$^{3,+}$,  
	 Christy Simpson$^{2}$, Thomas Roger$^{4}$, Jonathan S.\ Ben-Benjamin$^{1}$,  Daniele Faccio$^{2,}$}
	 
\email{+ These authors have contributed equally to this work.	 Corresponding author: daniele.faccio@glasgow.ac.uk}

\affiliation{ $^{1}$Texas A\&M University, College Station, Texas 77843-4242, USA\\
$^{2}$School of Physics \& Astronomy. University of Glasgow, Glasgow G12 8QQ UK \\
	$^{3}$Centre for Disruptive Photonic Technologies, TPI, Nanyang Technological University, 637371, Singapore\\
	$^{4}$Institute of Photonics and Quantum Sciences, Heriot-Watt University, EH14 4AS, UK
    }

\begin{abstract}
	
{Plasmonics and metamaterials have recently been shown to allow the control and interaction with non-classical states of light, a rather counterintuitive finding given the high losses typically encountered in these systems.  Here, we demonstrate a range of functionalities that are allowed with correlated and entangled photons that are used to illuminate multiple, overlaid patterns on plasmonic metasurfaces. Correlated photons allow to nonlocally determine the pattern that is imaged or, alternatively to un-scramble an image that is otherwise blurred. Entangled photons allow a more important functionality whereby the images imprinted on the metasurface are individually visible only when illuminated with one of the entangled photons.  Correlated single photon imaging of functional metasurfaces could therefore promise advances towards the use of nanostructured subwavelength thin devices in quantum information protocols.}\\
%
\\
Keywords: Plasmonic metasurfaces, Correlated imaging, Quantum imaging, Polarisation measurements

\end{abstract}

\date{\today}
\maketitle

Metamaterials and in particular metasurfaces have recently started to emerge as a platform that is viable for quantum processing at the single photon level. The first pioneering works demonstrated that quantum entanglement could be preserved in transmission through a metasurface \cite{24}, followed by evidence that photon indistinguishability could be preserved in passing from photons to plasmons, thus allowing to perform simple quantum processing steps such as Hong-Ou-Mandel bunching experiments directly on plasmonic chip \cite{25}. Recent experiments have also highlighted how the losses associated with metasurfaces may be harnessed as a resource \cite{26,27} to thus control the transmitted photon statistics \cite{28,29,30}.\\
Recent advances in metasurface optical design have provided ultra-thin devices that are capable of controlling and shaping the optical properties of a light beam, for example polarisation, orbital angular momentum (OAM) and focusing. More complex devices are also possible whereby the output depends on the input properties, for example the output OAM or an output holographic image can be controlled by varying the input polarisation \cite{31,32,33}.\\  
In the following, we will rely on measurements of nonlocal correlation properties of photons interacting with metasurfaces. Correlation detection with photon pair sources is an essential technique in experimental quantum optics {as it allows one to identify photons of interest to a particular experiment via their detection statistics}. This has enabled experimental results that serve as the foundation for quantum optics, for example the observation of the nonlocality of polarisation entangled photons \cite{1,2,3,4,5,6}, the Hong-Ou-Mandel two photon interference effect \cite{7,8,9}, nonlocal two-photon interference \cite{10,11}, dispersion cancellation \cite{12,13}, and the delayed choice quantum eraser \cite{14,15,16}. Further to that, technological developments based on correlation detection have provided more advanced imaging frameworks for studying the phenomenon of ghost imaging \cite{17,18,19,20} and polarisation, position and momentum characteristics of single photons \cite{21,22,23}.\\
In this work, we employ correlated and entangled photon pairs to demonstrate imaging capabilities whereby images become visible only as a result of the correlation or entanglement properties of the light used to illuminate the metasurface. To this end, we design metasurfaces that respond differently (provide different output images) depending on the input polarisation. In a first set of measurements we use a mixed (i.e. not entangled) photon state  to demonstrate nonlocal control of the projected images in the single photon regime by controlling the properties of a distant, correlated photon. 
This may be used to either nonlocally select an image or alternatively to nonlocally scramble (or unscramble) an image. \\
 \begin{figure*}[!t]
 	\centering 
 	\includegraphics[width=14cm]{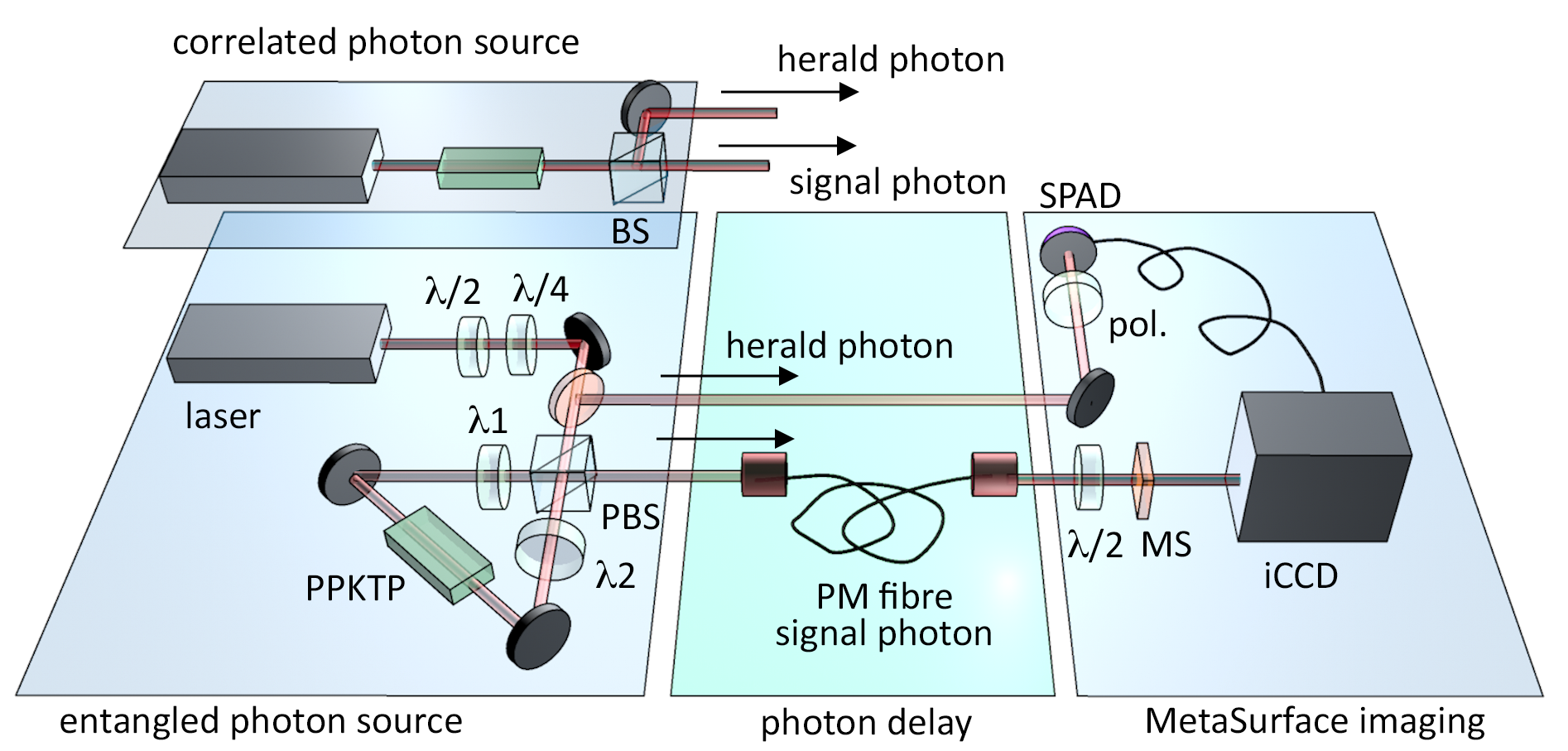}
 	\caption{ Polarisation heralded imaging with metasurfaces. Polarisation correlated photon pairs are generated by Spontaneous Parametric Down Conversion (laser at 404 nm pumping a PPKTP nonlinear crystal). To generate a mixed state, photons reflected at the beamsplitter (BS), form the `heralding' photons that are detected after polariser (pol.) that allows to select $H$ or $V$ heralding.  The transmitted photon of the pair,  the signal, is transmitted through the  polarisation-sensitive plasmonic metasurface (MS) and detected on an iCCD camera. Before the metasurface, we insert an fibre optical delay line so that the photon arrives on the iCCD  when the the camera electronic shutter is activated by the herald photon trigger. To generate an entangled state a more complex source is used based on a Sagnac interferometer with counterpropagating pump beams (controlled by quarter and half-wave plates). Two waveplates ($\lambda1$ and $\lambda2$) rotate the polarisations of the pump and SPDC photons respectively for one direction around the Sagnac loop. At the polarising beamsplitter (PBS), the entangled photons are directed, as before, to the herald and signal arm. A half-wave plate placed in front of the metasurface is used to rotate the photon polarisation state by 45 deg. that is equivalent to rotating the metasurface by 45 deg. \label{F:setup}}
 \end{figure*}
In a second set of experiments we use a source of pure-state, entangled photons and demonstrate how the metasurface can be used to {produce images} that are separately visible only in the presence of entanglement. Degrading the photon-pair entanglement subsequently degrades the image.\\
%
%
{\bf{Experiment: }} The experimental setup is shown in Fig.~\ref{1}. We generated pairs of photons with orthogonal polarisations at a wavelength of 808 nm by Spontaneous Parametric Down-Conversion (SPDC) in a type-II PPKTP nonlinear crystal that was pumped by a continuous-wave 100 mW laser at 404 nm wavelength. For the correlated (mixed state) photon measurements, {the photon pairs were  split with 50\% efficiency by a (non-polarising) 50:50 beamsplitter (BS). One of the pair, which we label ``\emph{herald}'', was detected with a Single Photon Avalanche Diode (SPAD), the output of which was connected to the external trigger of an intensified-CCD camera (iCCD, ANDOR iStar) and thus heralds the arrival of a photon at the camera sensor. The second photon of the pair, which we label ``\emph{signal}'', was optically delayed by 40 m of optical fiber (in order to compensate for the electronic delay acquired by the iCCD camera between the trigger arrival and the actual acquisition on the iCCD sensor) before being focused onto the metasurface sample and imaged onto the iCCD sensor by a pair of $\times$10 objective lenses (not shown for simplicity in Fig.~\ref{1}).\\
For the entangled measurements, the source of entangled photons is a counter-propagating Sagnac interferometer enclosing the PPKTP crystal \cite{34,35}. The input frequency-doubled 404 nm pump laser beam polarisation is fully controlled by $\lambda/4$ and $\lambda/2$ waveplates and is split into counterpropagating beams at the polarisation beam-splitter (PBS). {Two wavelength waveplates rotate the polarisation of the pump, without affecting the SPDC photon polarisation (indicated as $\lambda1$) and of the SPDC photons, without affecting the pump polarisation (indicated as $\lambda2$). The Sagnac interferometer thus produces an entangled output state from the PBS of the form $|H_h {V_s}\rangle+|V_h {H_s}\rangle$. }\\
\noindent 
{\bf{Mixed-state imaging of metasurface holograms.}}
The SPDC source creates both horizontally and vertically polarised herald and signal photons with a mixed state after the beamsplitter, with a density matrix operator:
\begin{eqnarray}\label{1}
 \hat{\rho}
& =&
\frac12
\left( \Big.
   \ket{H_h V_s} \bra{H_h V_s}
   +
   \ket{V_h H_s} \bra{V_h H_s}
\right)
\end{eqnarray}
 where indices $h$ and $s$ denote the {herald} and signal photons, respectively and $|H\rangle \langle H| $ and $|V\rangle \langle V|$ denote the horizontal and vertical polarisation states, respectively.
We note that we have neglected two other terms in the mixed state,
$\ket{H_h V_h} \bra{H_h V_h}$
and
$\ket{H_s V_s} \bra{H_s V_s}$
{{which do not contribute to the sub-ensemble of coincidence counts.}} 
\\
 In our experiments, we place a metasurface array in the optical path of the signal photons, with an operator described as 
 \begin{equation}
 \hat{M}=\vartheta_V (x,y) \hat\chi_s(0)+ \vartheta_H (x,y) \hat\chi_s(90^\circ)
 \end{equation}
where
$\hat \chi_s(\xi)$ is the polarisation projection operator (see SM)
for the signal photons,
$\Outer{\xi_s}{\xi_s}$,
along the polarisation angle $\xi$,
and where
$\vartheta_V (x,y)$ and $\vartheta_H (x,y)$ are the spatially $(x,y)$ and polarisation-dependent {transmission amplitude coefficients} of the metasurface. 
These amplitudes can be nonlocally controlled by the measurement process on the `herald' arm of the experiment. {The expectation value of the final measurement (i.e. the image that is observed on the iCCD camera) is given by $\langle O \rangle=\mbox{Tr}(\hat{\rho} \hat{\chi}_h (\phi) \hat{M}  (\xi) )$, where $\phi$ is the herald photon polariser angle (see SM). In the specific case in which we place an $H$ or $V$ polariser on the `herald' detector, the herald-arm polarisation operator $\hat{\chi}_h$, simply projects the photon state on to $\frac12 |V_s\rangle\langle V_s|$ or $\frac12 |H_s\rangle\langle H_s|$, respectively}.
 %
%
 %
 \\
Figure~\ref{F:results}(a), shows the detail of the first metasurface used that produces a star or a triangle, for $H_S$ or $V_S$ polarisation, respectively. This is obtained by superimposing horizontal and vertical slit antennas {[and by shifting the vertical slit antenna array from the horizontal slit antenna array by half the period {(see Methods for further details)]}}. The SEM image allows to identify the two distinct geometrical shapes, a star and a triangle. Here, our star metasurface is entirely composed of horizontal slit antennas:  selection of $H$  polarisation on the `herald' photon implies 
{{a vertical signal photon,}}
and thus produces a triangle image $\langle O_\blacktriangle \rangle = \frac{1}{2}\vartheta_V(x,y)$. By similar reasoning, selection of the $V$ polarisation on the herald photon produces a star image in the signal arm  $\langle O_\bigstar \rangle = \frac{1}{2}\vartheta_H(x,y)$
(where we assumed that the photon amplitude is uniform across the metasurface).
\begin{figure}[t!]
	\centering
	\includegraphics[width=7.8cm]{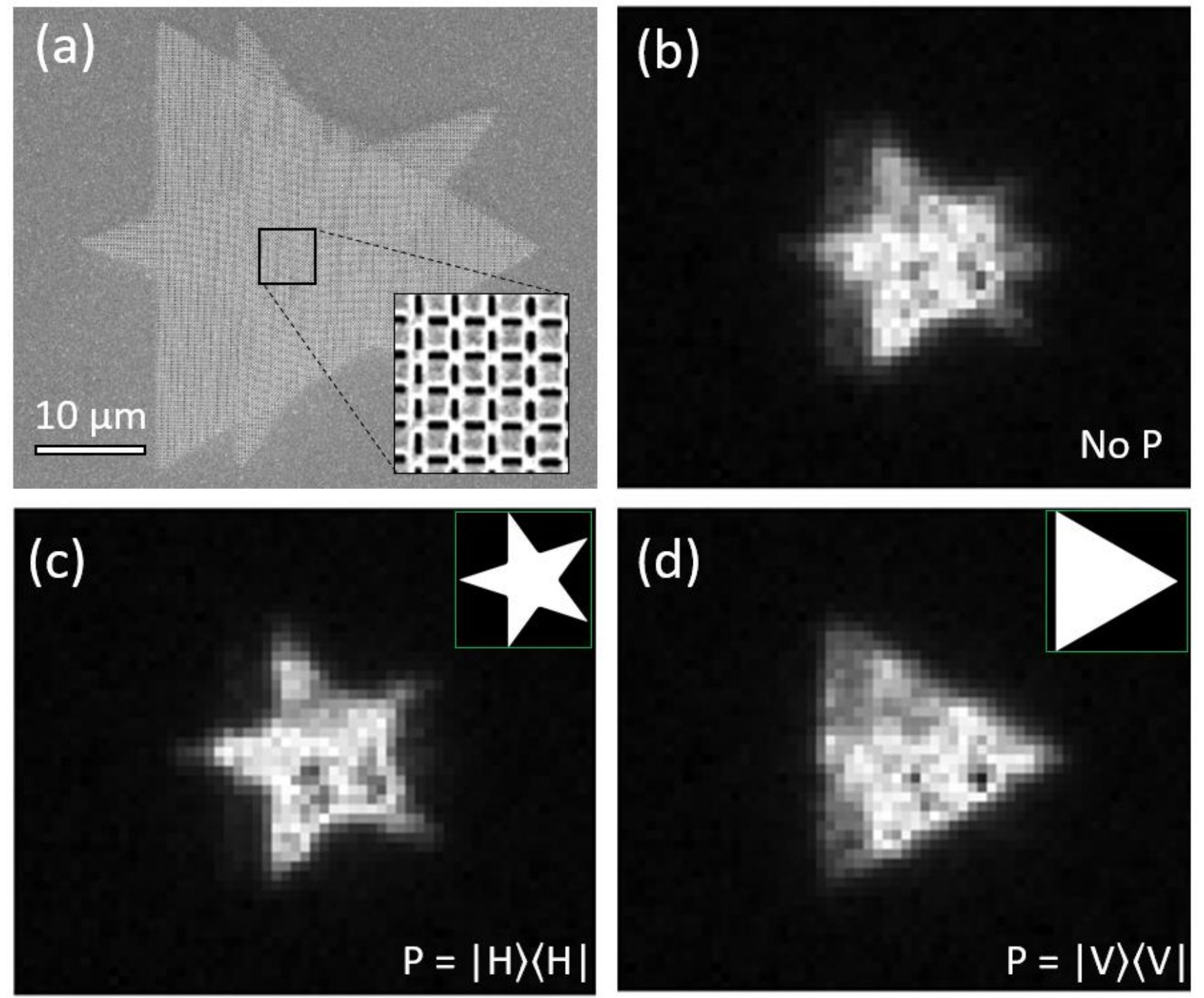}
	\caption{Heralded photon-control of  metasurface images. (a) The SEM image of the nanostructured metasurface is composed of horizontal and vertical slits making up two images: a star and a triangle. (b) If there is no polariser in the heralded optical path both shapes will be imaged. (c) By detecting only horizontally polarised heralded photons, the image of a star is recorded while, (d) the image changes to a triangle for the detection of vertically polarised heralded photons.  \label{F:results}}
\end{figure}
Figures~\ref{F:results}(c) and (d) show the images obtained {[compared with the case where no polariser is placed in the heralding optical path, Fig.~\ref{F:results}(b)]} confirming that only the star or the triangle is imaged depending on the selection of the herald photon polarisation.  \\
A second type of metasurface follows a more complex design. Contrary to the previously used metasurface, this nanostructured array modulates amplitude and also phase by using slit antennas fabricated in a multitude of lengths and of angles \cite{36,37,38}. \\
\begin{figure}[t!]
	\centering
	\includegraphics[width=8cm]{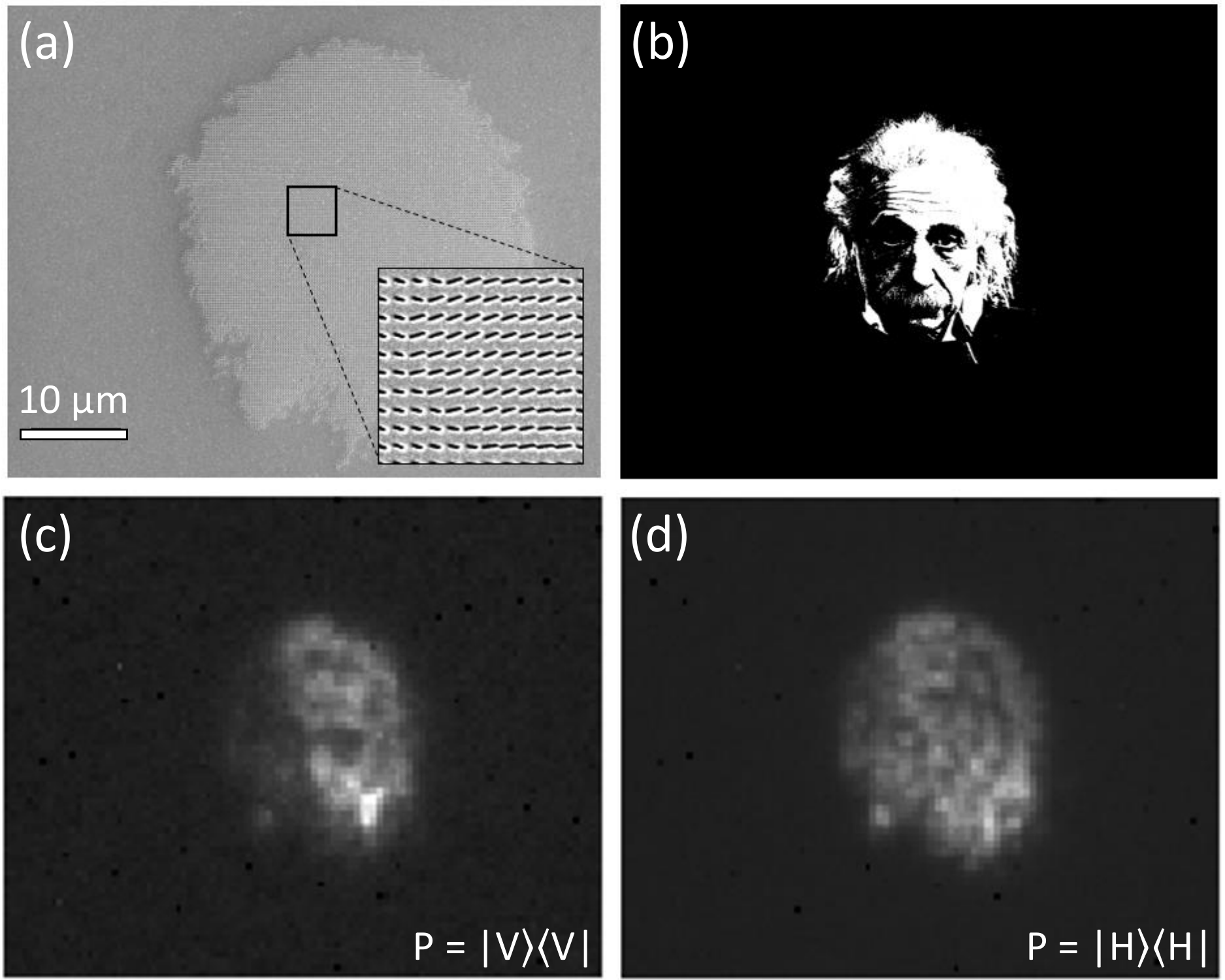}
	\caption{Hologram scrambling. (a) The SEM image of the nanostructured surface in is composed of slits of different lengths as to create different phases. The hologram was designed to generate the image of Albert Einstein's portrait, shown in (b). (c) By detecting only in the presence of $V$ polarised heralded photons, a clear image of the hologram is recorded, while  in (d) the image is scrambled in correspondence to $H$ polarised herald photons.   \label{F:results2}}
\end{figure}
Our fabricated metasurface, {shown in Fig.~\ref{F:results2}(a)}, has been designed to generate the portrait of Albert Einstein [Fig.~\ref{F:results2}(b)] for an incident horizontal polarisation. Thus by selecting $V$ polarised `herald' photons, we produce a clear hologram image of Albert Einstein's portrait as shown in Fig.~\ref{F:results2}(c). Alternatively, when we image in correspondence to $H$ polarised `herald' photons, the hologram image is scrambled, as shown in Fig.~\ref{F:results2}(d). The image is lost in this case due to having the orthogonal polarisation incident which will behave very differently for the angled slit antennas, thus affecting both the modulated amplitude and the modulated phase. \\
{\bf{Pure-state imaging of metasurface {structures.}}}\\
Using the experimental layout shown in the lower left panel of Fig.~\ref{F:setup} we generate an entangled state $|\Psi\rangle=(|H_h {V_s}\rangle+|V_h {H_s}\rangle)/\sqrt 2$ with a density matrix operator $ \hat{\rho}={|\Psi\rangle \langle \Psi|}$. We now also arrange the the setup so that the star-triangle metasurface shown in Fig.~\ref{F:results}(a) is placed at 45 deg.  with respect to the polarisation axis of the photons. The metasurface operator can now be written as
 \begin{eqnarray}
 \hat{M}
=
\frac12
\vartheta_D (x,y)
\hat\chi_s(45^\circ)
+
\frac12
\vartheta_{AD} (x,y)
\hat\chi_s(135^\circ)
%
 \end{eqnarray}
 where $D$ and $AD$ subscripts indicate diagonal and anti-diagonal polarisation terms.
%
Calculation of the expectation value $\langle O \rangle=\mbox{Tr}(\hat{\rho} \hat{\chi}_h (\phi) \hat{M} { (\xi)})$ reveals that for a mixed state, we will always see a superposition of both the {polarisation dependent patterns}, i.e. a superposition of a star and a triangle. However, in the presence of a pure state, imaging only in the presence of a $D$ (or $AD$) herald photons will selectively image only the $AD$ (or $D$) metasurface pattern, i.e. the star or triangle alone will become clearly visible without any overlap of the other.\\
 \begin{figure}[t!]
	\centering
	\includegraphics[width=8cm]{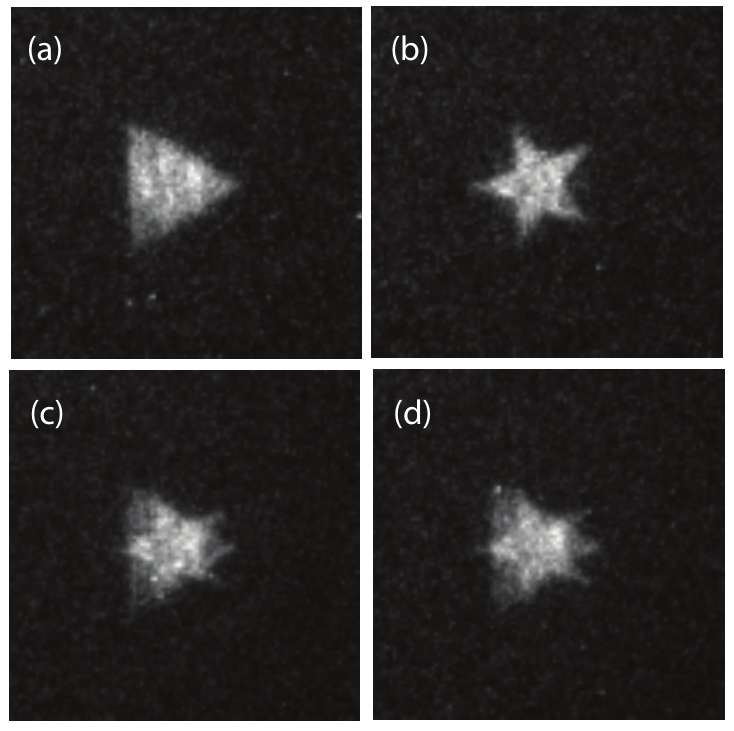}
	\caption{Imaging with entangled photons:  Images obtained with entangled states (measured Bell parameter $S=2.5$) with the herald polariser selecting photons at (a) 45 deg and (b) 135 deg. The same images obtained with a mixed state (measured Bell parameter $S=1.6$) at (c) 45 deg and (d) 135 deg. \label{F:results3}}
\end{figure}
 Figures~\ref{F:results3}(a) and (b) show the experimental measurements obtained for entangled photons, when selecting $D$ and $AD$ herald photons, respectively. We separately measured the Bell parameter for the photon state used in this experiment to be $S=2.5$ (i.e. above the threshold $S=2$ for entanglement): the triangle and star are individually very clearly visible, with high contrast and no visible contribution of the other shape. {The Sagnac interferometer can also be used to produce a mixed polarisation state of the form of \eqref{1} by rotating the $\lambda2$ to 0 deg such that the PPKTP crystal is pumped in both directions around the Sagnac loop but there is no compensation for the SPDC-photon temporal walk-off  occurring within the crystal. By doing so, the experiment can be repeated with non-entangled photons with a Bell parameter that was measured to be $S=1.6$.} The results are shown in Figs.~\ref{F:results3}(c) and (d), that look nearly identical regardless of the herald photon polarisation and show a clear superposition of both the star and triangle. \\
 %
  \begin{figure}[t!]
	\centering
	\includegraphics[width=8.5cm]{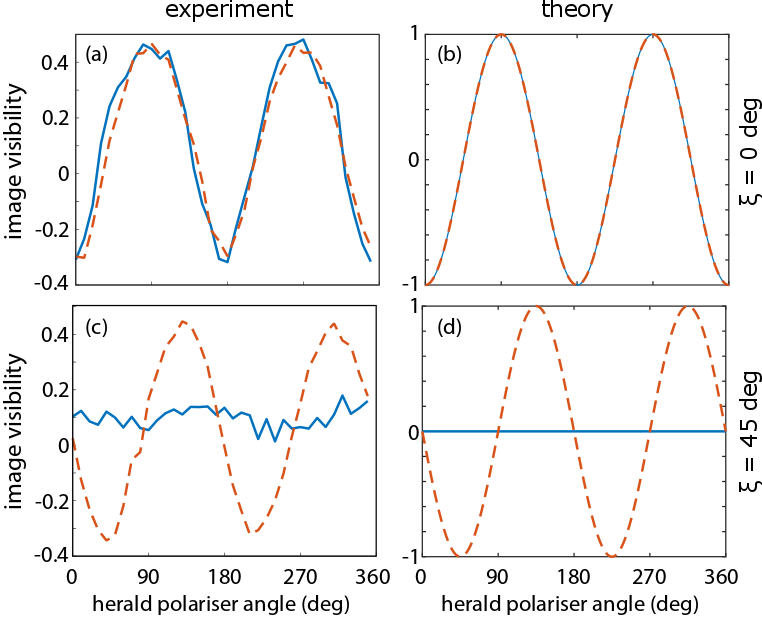}
	\caption{Imaging with entangled photons:  Image visibility,  $V=(O_\blacktriangle-O_\bigstar)/(O_\blacktriangle+O_\bigstar)$, for the `triangle' image plotted versus the herald photon polariser angle. Experimental (a) and theoretical (b) results for the case of the metasurface aligned along the H-V axis of the input photons. Experimental (c) and theoretical (d) results for the case of the metasurface aligned at 45 deg with respect to the polarisation of the input photons. In all figures, solid lines refer to an input mixed state and dashed lines refer to input pure states with measured Bell parameter $S=2.5$. \label{F:results4}}
\end{figure}
Figure~\ref{F:results4} shows the full results for these measurements (i.e. for varying angles of the selected herald photon polarisation from 0 to 360 deg), with a direct comparison to the theoretical predictions. In particular, we calculate and measure  the visibility of just the `triangle' image, $V=(O_\blacktriangle-O_\bigstar)/(O_\blacktriangle+O_\bigstar)$. For the case when the metasurface is aligned with the H-V axis of the input photons ($\xi=0^\circ$), $V_{pure} = V_{mixed} = - \cos(2\phi)$. Alternatively, for the more interesting case in which the metasurface angle $\xi=45^\circ$, we predict
\begin{equation}
V_{\textrm{mixed}}= 0 \, \, \textrm{and} \, \, V_{\textrm{pure}} = -\sin(2\phi)
\end{equation}
for the mixed and pure states, respectively.
As can be seen in Fig.~\ref{F:results4}, there is a good agreement between the experiment and theory, although the visibility is lower in the experiment due to background noise on the iCCD sensor. Nevertheless, the main features are clearly observable. Figures~\ref{F:results4}(a) and (b) show the case in which the metasurface is aligned parallel to the H-V polarisation of the photons: the image observed intensities are essentially identical for the cases of input mixed and pure states, i.e. there is no discernible advantage or difference using entangled states. However,  Figs.~\ref{F:results4}(c) and (d) show the case in which the metasurface is aligned at 45 deg to the H-V polarisation of the photons: now the mixed state shows zero visibility whereas imaging with entangled photons gives rise to clear oscillations in the `triangle' visibility. Each peak corresponds to all photons in the `triangle' image and none in the `star' image and each trough corresponds to the opposite situation. With these settings, the images are visible only when using entangled photons.\\
{\bf{Conclusions.}}
By making a polarisation measurement of a heralding photon, two effects are explored by using two different input states. With input mixed states, image selection is achieved through metasurfaces constructed solely of horizontally and vertically angled slit antennas. By heralding with either horizontally or vertically polarised photons, one and only one of the two metasurface {patterns} is imaged. In addition, a scrambling effect is presented by using a metasurface hologram that modulates both phase and amplitude. In this instance, the detection of the vertically polarised `herald' photons results in an image of the hologram, and the detection of the horizontally polarised `herald' photons will generate a noisy, scrambled image. \\
{With input states that are entangled, we show that under the assumption of only $H$ and $V$ photon illumination, it is possible to clearly distinguish the individual {images imprinted on the  metasurface} i.e. individual {images} become visible only in the presence of pure, entangled states. 
This functionality is the result of quantum interference occurring on the metasurface, in line with recent reports of `quantum metamaterials' \cite{moti}.}\\
 The wavelength dependence of metasurfaces may create further opportunities for encrypting sequences of images at different wavelengths for single photon communication channels and the diversity of metasurface designs also opens up the possibility of spatially multiplexed imaging systems which, when combined with time-resolved imaging, can be used for quantum state tomography and exploration of entangled states with  imaging techniques \cite{39}.\\
\\

All experimental data is available at \\

{\textit{Acknowledgements.}} The authors thank Prof Nikolay Zheludev for fruitful discussions. The authors acknowledge the support of the  the support of the Singapore MOE Grants MOE2011-T3-1-005 and MOE2016-T3-1-006, ASTAR QTE Programme Grant SERC A1685b0005, EPSRC (U.K.) grants EP/M009122/1 and EP/J00443X/1 and EU Grant ERC GA 306559. CA acknowledges Robert A. Welch Foundation (Grant No. A-1261) and the Bio-Photonics initiative of the Texas A\&M University.\\

{\bf{Methods.}}\\

{\bf{Working principle and design procedure of metasurfaces}}. We use uniform slit antennas for designing the star-triangle metasurface. The period is 300 nm by 300 nm, and the slit dimension is 190 nm by 50 nm. At the TM polarization (electric field is perpendicular to the long axis of the slit), the incident wave at a wavelength of 808 nm can efficiently
transmit through the subwavelength slit through plasmonic resonance. The simulated and experimental transmittance is $60\%$ and $36\%$ respectively. To multiplex two images which are only responsive to either H or V polarization, we fill up the image with the slit antennas with orthogonal orientation. These two sets of slits are laterally shifted by a half period
in order to reduce their mutual coupling and thus minimize the crosstalk between the two images.\\
The design principle of the Einstein metasurface relies on the Babinet plasmonic metasurface. In order to generate a clear holographic image at a certain propagation distance, we use nonuniform slit antennas with varying geometric parameters (length and orientation) to simultaneously control both the transmission amplitude (continuous) and phase (8 levels). The
first four types of slits have the same width of 50 nm and different lengths (170 nm, 200 nm, 240 nm, and 280 nm), and their long axis is orientated in the second and fourth quadrants, creating a step phase delay of 0, $\pi$/4, $\pi$/2 and 3$\pi$/4. The second four types of slits have the same geometric parameters as the first set but their long axis is
orientated in the first and third quadrants in order to create additional π phase delay to cover the phase of $\pi$, 5$\pi$/4, 3$\pi$/2 and 7$\pi$/4. The amplitude control in the orthogonal polarization $|V\rangle$ to the incident linear polarization $|H\rangle$ follows a simple $\sin2\theta$
rule where $\theta$ is the angle of the slit normal with respect to horizontal axis.
\\
\textbf{Fabrication of metasurfaces.}
The metasurfaces are fabricated in a 100 nm-thick gold film with focused ion beam milling (FEI Helios 650, 30 kV, 7.7 pA). The film is deposited on a silica glass substrate by using thermal evaporation at a rate of 0.2 Angstroms/second (Oerlikon 250, Germany) after being properly cleaned by acetone and isopropanol. The CAD files are created from desirable pixilated amplitude and phase profiles using Matlab coding.\\


\newpage
\onecolumngrid


\section{Supplementary Information: Quantum metasurface theory.}
%
%
%

In our experiment,
we produce photon pairs
in two states;
a mixed state
and
a pure state.
The mixed state $\hat \rho_{\subsc{mixed}}$
\begin{align}
\hat \rho_{\subsc{mixed}}
&=
\frac12 \Outer{H_h V_s}{H_h V_s} + \frac12 \Outer{V_h H_s}{V_h H_s}
\end{align}
with classical probabilities $\frac12$
for the two terms,
and the pure state $\hat \rho_{\subsc{pure}}$
\begin{align}
\hat \rho_{\subsc{pure}}
&=
\frac12
\left( \Bigg.
   \ket{H_h V_s}
   -
   \ket{V_h H_s}
\right)
\left( \Bigg.
   \bra{H_h V_s}
   -
   \bra{V_h H_s}
\right)
\\&=
\frac12
\left( \Bigg.
   \Outer{H_h V_s}{H_h V_s}
   -
   \Outer{H_h V_s}{V_h H_s}
   -
   \Outer{V_h H_s}{H_h V_s}
   +
   \Outer{V_h H_s}{V_h H_s}
\right)
.
\end{align}

To herald an H-photon
with a polariser at some angle $\phi$,
we perform a partial trace over the herald photon of
the density matrix with the polarisation projection operator $\hat \chi$
\begin{align}
\hat \chi_h(\phi)
&=
\Outer{\phi}{\phi}
\\&=
\left( \Bigg.
   \cos\phi
   \ket{V_h}
   +
   \sin\phi
   \ket{H_h}
\right)
\left( \Bigg.
   \cos\phi
   \bra{V_h}
   +
   \sin\phi
   \bra{H_h}
\right)
\\&=
\cos^2\phi
\ket{V_h}
\bra{V_h}
+
\cos\phi
\sin\phi
\ket{V_h}
\bra{H_h}
+
\cos\phi
\sin\phi
\ket{H_h}
\bra{V_h}
+
\sin^2\phi
\ket{H_h}
\bra{H_h}
.
\end{align}
%

\bigskip
\noindent%
\textbf{Heralding a photon.}
To see the (un-normalised) state $\hat\rho{(s)}$ of the signal photon
given a measurement of the herald photon%
we perform a partial trace
on the heralded photons
with the herald polariser operator.
After heralding
a photon
through a polariser at some angle $\phi$,
the quantum state
impinging on the metasurface becomes
(for our two states, $\hat \rho_{\subsc{mixed}}$ and $\hat \rho_{\subsc{pure}}$) 
\begin{align}
\hat \rho_{\subsc{mixed}}^{(s)}
&=
\Trace{h}{ \hat\chi_\phi \otimes \hat 1_s \, \hat \rho_{\subsc{mixed}} }
=
\frac12 \cos^2 \phi \Outer HH
+
\frac12 \sin^2 \phi \Outer VV
\\
\hat \rho_{\subsc{pure}}^{(s)}
&=
\Trace{h}{ \hat\chi_\phi \otimes \hat 1_s \, \hat \rho_{\subsc{pure}} }
=
%
\frac12
\left( \Bigg.
   \sin\phi
   \ket V
   -
   \cos\phi
   \ket H
\right)
\left( \Bigg.
   \sin\phi
   \bra V
   -
   \cos\phi
   \bra H
\right)
%
\end{align}
where we omitted the `s' subscripts 
since at this level
we only have signal photons.

\bigskip
\noindent%
\textbf{Passage through metasurface.}
The passage through the metasurface
oriented along the angle $\xi$
could be modeled with the operator
\begin{align}
\hat M
=
\vartheta_\blacktriangle(\xi)
\hat \chi_s(\xi)
+
\vartheta_\bigstar(\xi)
\hat \chi_s(\xi+90^\circ)
%
\end{align}
Considering only the $\blacktriangle$-part
(the star part will follow along the same lines),
we find that the photon intensity
passing through the metasurface is
(for our two states, mixed and pure)
\begin{alignat}{2}
&I_{\subsc{mixed}}^\blacktriangle
&&=
\vartheta_\blacktriangle (\xi)
\trace{ \hat \chi_\xi \, \hat \rho_{\subsc{mixed}}^{(s)} }
=
\frac12
\vartheta_\blacktriangle (\xi)
\left[
   \cos^2\phi \sin^2\xi
   +
   \sin^2\phi \cos^2\xi
\right]
\\
&I_{\subsc{pure}}^\blacktriangle
&&=
\vartheta_\blacktriangle (\xi)
\trace{ \hat \chi_\xi \, \hat \rho_{\subsc{pure}}^{(s)} }
\nonumber
\\&
&&=
\frac12
\vartheta_\blacktriangle (\xi)
\left[
   \sin^2\xi \cos^2\phi
   -
   2 \cos\xi \sin\xi \cos\phi \sin\phi
   +
   \cos^2\xi \sin^2\phi
\right]
%
\end{alignat}
Similarly,
the intensity of a signal photon
transmitted through a pixel in the $\bigstar$ region
of the metasurface is
\begin{alignat}{2}
&I^\bigstar_{\subsc{mixed}}
&&=
\frac12 \vartheta_\bigstar(\xi)
\left[
   \cos^2\phi\cos^2\xi+\sin^2\phi\sin^2\xi
\right]
\\
&I^\bigstar_{\subsc{pure}}
&&=
\frac12 \vartheta_\bigstar(\xi)
\left[
   \cos^2\phi\cos^2\xi
   +
   2\cos\phi\sin\phi\cos\xi\sin\xi
   +
   \sin^2\phi\sin^2\xi
\right]
.
%
\end{alignat}
Recall that the $\vartheta$'s
are functions of position.
So to define the visibility,
we integrate over the position and normalise the
total area (of both the $\bigstar$ and $\blacktriangle$) to unity.
The visibility is
\begin{align}
V = \frac
   {I^\blacktriangle-I^\bigstar}
   {I^\blacktriangle+I^\bigstar}
,
%
\end{align}
and using that
$I_{\subsc{mixed}}^\blacktriangle+I_{\subsc{mixed}}^\bigstar=1/2$
and that
$I_{\subsc{pure}}^\blacktriangle+I_{\subsc{pure}}^\bigstar=1/2$
we find that the visibilities are
\begin{align}
V_{\subsc{mixed}}
=
\left(
   2\cos^2\phi-1
\right)
\left(
   2\sin^2\xi-1
\right)
%
,
\end{align}
and
\begin{align}
V_{\subsc{pure}}
=
V_{\subsc{mixed}}
-
\sin(2\phi)
\sin(2\xi)
%
%
.
\end{align}
Placing the metasurface at $45^\circ$,
we find that the visibility of the mixed state
is constant (zero), and the visibility of the pure state is $-\sin(2\phi)$.

\end{document}